\documentclass[journal,twocolumn]{IEEEtran}
\usepackage{cite}
\usepackage[pdftex]{graphicx}
\usepackage{amsmath}
\usepackage{algorithmic}
\usepackage{xcolor}
\usepackage{array}
\usepackage{algorithm,algorithmic}
\usepackage[utf8]{inputenc}
\usepackage{enumitem} 
\usepackage[many]{tcolorbox}
\usepackage{epsfig} 
\usepackage{pdfpages} 
\usepackage{amsmath, amssymb, mathrsfs}
\usepackage{epstopdf}

\usepackage{color}
\usepackage{subfig}

\usepackage{cite}
\usepackage{tikz}
\usepackage{pgfplots,pgfplotstable, booktabs}
\usetikzlibrary{pgfplots.groupplots, matrix,backgrounds}
\usepgfplotslibrary{fillbetween}
\usepackage{filecontents}
\usepackage{graphicx}
\usepackage{float}
\usepackage{subfig}
\usepackage{tikz}
\usepackage{pgfplots,pgfplotstable, booktabs}
\usetikzlibrary{shapes,arrows}
\usetikzlibrary{positioning, fit, calc}  \usetikzlibrary{arrows}
\usepackage{xcolor}  
\tikzset{block/.style={draw, thick, text width=2cm , minimum height=1.3cm, align=center},   
line/.style={-latex}     
} 

\usepackage{graphicx}
\usepackage{pgfplots}
\pgfplotsset{width=0.5\columnwidth,compat=1.17}

\pgfplotsset{
  every axis plot/.append style={line width=0.8pt},
  every axis plot post/.append style={
  }
}

\newcommand\mydots{\hbox to 1em{.\hss.\hss.}}

\begin{document}

\title{Model Predictive Control for a Medium-head Hydropower Plant Hybridized with Battery Energy Storage to Reduce Penstock Fatigue}


\author{Stefano~Cassano
        and~Fabrizio~Sossan% <-this % stops a space
\thanks{S. Cassano and F. Sossan are with the Centre for processes, renewable energies and energy systems (PERSEE) of MINES ParisTech, Sophia Antipolis, France. E-mail: {stefano.cassano, fabrizio.sossan}@mines-paristech.fr.}% <-this % stops a space
\thanks{Research supported by the European Union Horizon 2020 research and innovation program under XFLEX HYDRO, grant agreement 857832.
%of the Hydropower Extending Power System Flexibility project (XFLEX HYDRO, grant agreement No 857832).
}
\thanks{This paper has been submitted to PSCC 2022.}
}

\maketitle

\begin{abstract}
A hybrid hydropower power plant is a conventional HydroPower Plant (HPP) augmented with a Battery Energy Storage System (BESS) to decrease the wear and tear of sensitive mechanical components and improve the reliability and regulation performance of the overall plant. A central task of controlling hybrid power plants is determining how the total power set-point should be split between the BESS and the hybridized unit (power set-point splitting) as a function of the operational objectives. This paper describes a Model Predictive Control (MPC) framework for hybrid medium- and high-head plants to determine the power set-point of the hydropower unit and the BESS. The splitting policy relies on an explicit formulation of the mechanical loads incurred by the HPP's penstock, which can be damaged due to fatigue when providing regulation services to the grid. By filtering out from the HPP's power set-point the components conducive to excess penstock fatigue and properly controlling the BESS, the proposed MPC is able to maintain the same level of regulation performance while significantly decreasing damages to the hydraulic conduits. A proof-of-concept by simulations is provided considering a 230 MW medium-head hydropower plant.

\end{abstract}


% Use this to place sponsorships

\section{Introduction}
Hydropower is a key asset of the electrical power system infrastructure, providing both a significant amount of electricity and regulation services to the power grid.
In 2018 hydropower accounted for 16.8\% of the total European power production and 70\% of all renewable generation \cite{globe}. 

The increasing proportion of stochastic generation in the electric power systems will require steeper ramping duties and more frequent start-and-stop's to all dispatchable generation units. Increased regulation duties for hydropower plants (HPPs) will determine increased wear and tear levels, ultimately leading to increased maintenance needs and negatively affecting plants' economics.  For example, in medium- and high-head HPPs, steep changes of the plant set-point aggravate water hammer effects and pose a serious risk of damaging the hydraulic conduits because of the fatigue levels incurred by the penstock, as demonstrated in \cite{Dreyer2019DigitalCF}. Similarly, in low-head (or run-of-the-river) HPPs, increased regulation duties determines wear and fatigue of the guide vane bushes and the Kaplan turbine's blade actuator (e.g., \cite{GERINI2021100538}). All these wear-and-tear effects are a serious concern for operators because their maintenance and replacement is very expensive and typically require to shut down the plant for long periods.

% whose maintenance is difficult as it normally involves removing the turbine from its hub and shutting down the plant for days or weeks.

A solution advocated in the technical literature to limit damages to mechanical components and increase the reliability and flexibility of an HPP is adding, in parallel to the hydraulic turbine, a battery energy storage system (BESS) to supply fast variations of the power output. This configuration is generally referred to as a hybrid power plant. Hybridization with a BESS can be applied to other kinds of resources too, typically, however, for improving dispatch performance. Hybrid power plants came to prominence due to decreasing prices of BESSs, especially Lithium-ion batteries that excel in supplying quick variations of power due to their fast kinetics, lack of mechanical time constants, and power-electronic interface. The principle underlying the notion of hybrid power plants is that operating the conventional plant and BESS within a unified control framework brings benefit over operating the two units separately, thanks to leveraging SCADA information or plant specificities that would not be otherwise available outside plant premises. This is particularly true for fatigue reduction, where required control actions are on a time scale of few seconds, and specific plant information, such as power set-points and speed governor properties, and measurements are needed.

\begin{figure}[!b]
\center
\includegraphics[width=1\columnwidth]{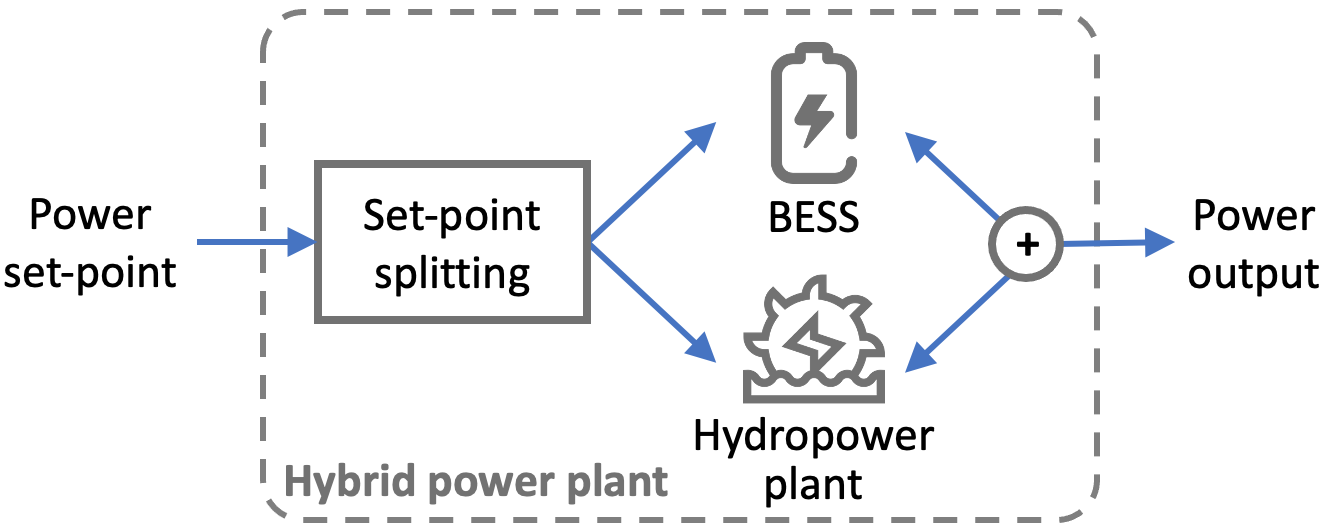}
\caption{A hybrid hydropower plant.}\label{fig:hybrid}
\end{figure}

When controlling a hybrid power plant, a central objective of the problem is to determine how the total power plant output is shared among the two (or possibly more) controllable resources. This situation is exemplified in Fig.~1: a single power set-point for the hybrid plant needs to be split into two, one for the conventional generation unit (e.g., turbine), and another for the BESS. We call this procedure power set-point splitting, or splitting.

%The problem tackled inof a hybrid HPP consists in designing the control set-point for the BESS and HPP such that the aggregated contribution of the two resources equals the prescribed total power output. In addition, the set-point of each resource should be such to minimize the wear and tear of the plant and the cycling of the battery.

Addressing meaningful splitting policies targeting fatigue reduction in HPPs is a relatively new research problem, which has been addressed in the literature with empirical models so far. For example, a commonly proposed strategy is low-pass filtering the power set-point; then, the filtered set-point is sent to the HPP for actuation, and the rest is taken by the BESS (\cite{7514942,6039893}). There are two main assumption underlying this approach. The first is that high-frequency variations of the power set-point are the root cause of fatigue in mechanical components; second, it assumes the existence of a formal procedure to translate from fatigue requirements to cut-off frequency of the filter. This procedure is, however, not documented in the existing literature. Both these assumptions are difficult to be justified and realized in practice. In particular, since the fatigue and mechanical loads acting on the components are not modeled explicitly, low-pass filtering might result in conservative estimates of the fatigue level, ultimately resulting in excess use of the BESS and not effective fatigue reduction. In addition, setting the cut-off frequency of the filter might result in approximations and oversimplification of wear-and-tear processes.

For the first time in the literature, this paper proposes a splitting strategy for a hybrid HPP that aims at explicitly modeling the impact of mechanical loads on fatigue. Unlike empirical methods that are typically uninformed of the actual fatigue levels of the mechanical components, the proposed method uses a formal way to model fatigue and thus tackle the power splitting problem in a measurable and quantifiable way.  This is thanks to using (linearized) models of the mechanical loads in the penstock and deriving a closed-form expressions to limit the mechanical stress below a critical level.

The rest of this paper is organized as follows. Section II illustrates the case study and problem, Section III describes the methodology of the proposed splitting policy, Section IV discusses simulation results on a detailed HPP model and impact of modeling approximations introduced for the MPC, and Section V concludes.

\section{Case study and problem statement}
The hybrid power plant controller proposed in this paper is specifically developed for medium- and high-head HPP, which have been documented reporting increased fatigue to the hydraulic pressurized conduits (penstock) when providing primary and secondary frequency regulation \cite{Dreyer2019DigitalCF}.

The principle of fatigue is that accumulated cycles of alternating mechanical loads (i.e., forces) on a component ultimately lead to ruptures and failure. For medium- and high-head HPPs, rapid fluctuations of the power set-point determine a pressure shock-wave in the conduits (water hammer), damaging the penstock in the long run. 

In this context, the central objective of hybridizing a medium- or high-head HPP with a BESS is 1) adjusting the set-point of the HPP to limit water hammer, thus preserving the penstock from fatigue and avoiding expensive maintenance and, 2), use the BESS to compensate for the missing regulation from the HPP.

An important element of the splitting policy proposed here is a model predictive control that, by modeling the forces acting on the penstock walls resulting from a change of set-point, removes those components of the HPP power set-point conducive to excess fatigue. Based on this information, the BESS set-point is then computed.  The MPC for fatigue reduction (proposed originally in \cite{Cassano2021}) is summarized in  \ref{sec:mpc} for the sake of clarity. Then, its extension with the integration of the BESS, which is the central contribution of this paper and the final essential component of the hybrid power plant controller, is described in \ref{sec:bess}.

This paper assumes, for simplicity, an HPP with a single hydraulic turbine and penstock. However, the methodology is general and could be extended to accommodate other setups provided that the specific plant models required by the MPC are available.

The main elements of an HPP and its speed governor are shown in Fig.~2, now briefly described for introducing the nomenclature used in the rest of the paper. The feedback loop on the rotor speed, $\omega_r$, implements a frequency droop controller, where $R$ is the droop coefficient, whose task is keeping the rotational speed of the machine near a design value, $\omega_0$. The PID controller determines the position of the so-called guide vane, indicated by $y^\star$, which controls the flow of water to the turbine. The parameters of the PID controller (that also includes a ramp limiter,  actuators' dynamics, and saturation limits,  not shown here for compactness) are typically chosen to obtain specific timing properties of the power output to comply with eligibility criteria for grid frequency regulation (see the Results section for an example). Finally, "Power reference" implements the so-called speed changer setting, which allows the operator to change the plant power set-point according to an external reference signal. As known, this is typically used to implement, e.g., secondary frequency regulation set-points, rescheduling, and electricity market commitments.

\begin{figure}
\center
\includegraphics[width=1\columnwidth]{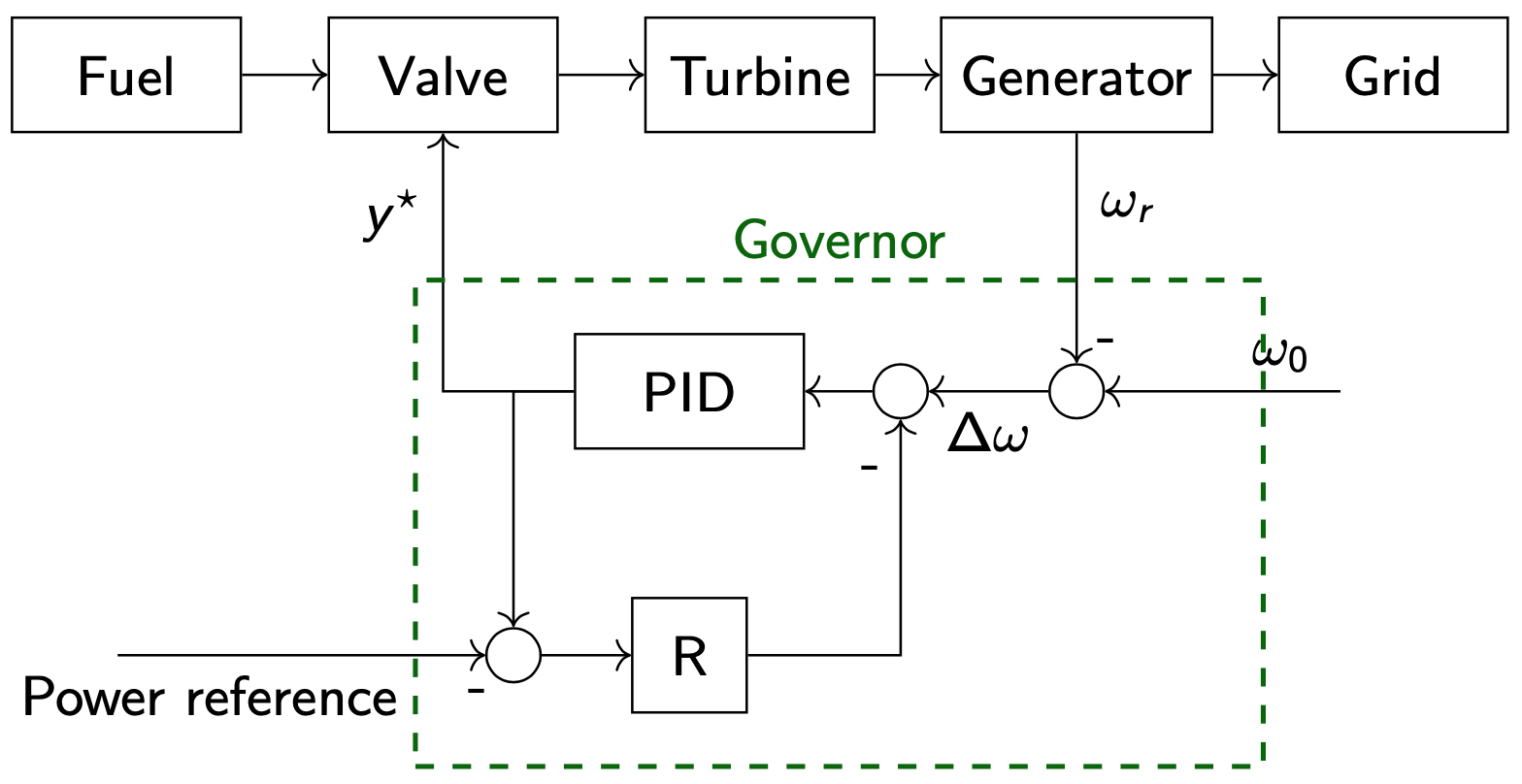}
\caption{Components and governor of a hydropower plant.}
\end{figure}

In the rest of this paper, the symbol $y^\star$ refers to the guide vane computed by the (standard) governor of Fig.~2. The action of $y^\star$, although tuned to respect the static design limits of the plant, is however uninformed about mechanical load dynamics in the penstock, and fatigue effects (\footnote{Stricter limits to reduce load dynamics' impact could be imposed by, e.g., altering the gains of the PID controller or changing the rate limiter. However, this might result in conservative decisions that limit regulation capacity. In addition, this strategy would still be trial and error, offering no formal methods to account for induced fatigue.}).

The objective of the hybrid power plant's controller proposed in this paper is to avoid those patterns of $y^\star$ conducive to premature aging of the penstock and use the BESS to complement the missing regulations. It is worth highlighting that the current paper focuses only on penstock aging, and not BESS aging. This is based on the assumption that the cost of maintaining (or replacing) a critical piece of infrastructure like the penstock is much larger than aging a BESS. This hypothesis is preliminarily corroborated by the findings of this paper, which shows that the BESS power rating and energy capacity required to limit penstock fatigue is a small fraction of the plant's rated power; thus, BESS costs might be small. Aspects related to cycle aging of the BESS, which are however of definite general interest, will be investigated in future research to verify how the control action can also impact minimally on BESS residual life, as done, e.g., in \cite{xu2014bess, namor2016assessment} in other contexts.

\section{Methods}

\subsection{Modeling of hydropower plants}\label{sec:hppmodels}
HPP models for power systems studies are, typically, transfer function and equivalent circuit models. They are also referred to as one-dimensional (1-D) models, as opposed to computational fluid dynamic (CFD) models in 2- or 3-D, typically used to simulate the detailed behavior of a single hydraulic component and generally not suited to power systems simulations due to their computational complexity.

\begin{figure}
\includegraphics[width = 1\columnwidth]{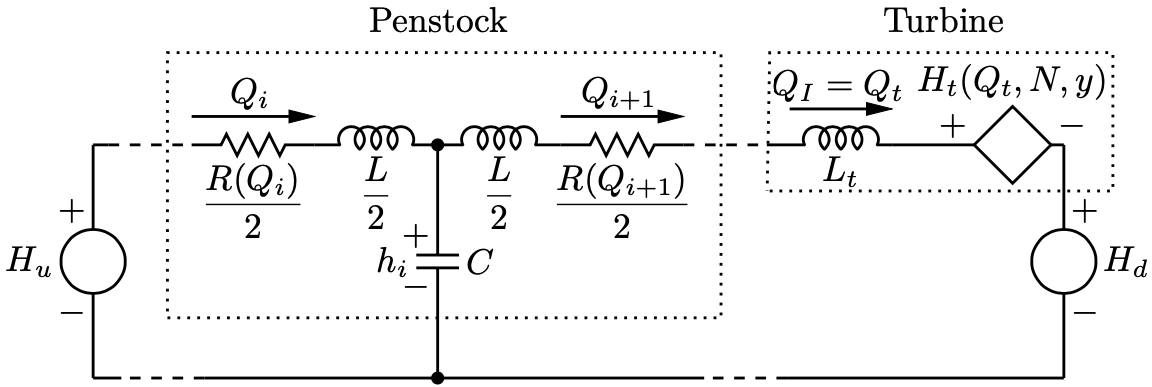}  \caption{Continuous-time equivalent circuit model of a medium-head hydropower plant. }\label{fig:EEC}
\end{figure}

The equivalent circuit analogy consists in modeling the piezometric head (or head, which refers to the dynamic and static pressure in terms of water column, in meters), denoted by $H$, and water discharge, $Q$, at a given point of a hydraulic circuit as a voltage and current of an electrical circuit \cite{electric}. The equivalent circuit model, which will be now described, of a medium-head HPP is shown in Fig.~\ref{fig:EEC}.
High-head power plants typically also included a surge tank, modeled however with the same equivalent circuit principles.

The equivalent circuit model in Fig.~\ref{fig:EEC} describes the flow of water and the pressures within the hydraulic circuit of the plant: the two voltage sources at the circuit's far ends, $H_u$ and $H_d$, are the water height of the upstream and downstream reservoir, respectively; then, the series of RLC circuits model the penstock, and the controlled voltage source $H_t$ models the turbine. Penstock's and turbine's models are now summarized.

The penstock is described by the momentum and mass conservation laws of water applied to the penstock. Its model corresponds to a set of hyperbolic partial differential equations (PDEs). A solution to solve it comes from the field of telecommunication with the telegraphist's equation (\cite{evans_partial_1998}), which can be solved with numerical methods. This is conventionally done by discretizing the penstock conduit in a finite number of elements, with each element modeled as a third-order RLC circuit, as shown in Fig.~\ref{fig:EEC} for penstock element $i$: the voltage $h_i$ is the water head (or static pressure) in the central part of the penstock element; $Q_i$ and $Q_{i+1}$ are the water flows at the receiving and sending ends. This model captures head losses along the conduit, and most importantly, pressure dynamics within the penstock, which are at the origin of the water hammer effect. The circuit parameters of the penstock model can be determined based on penstock physical properties, as described in \cite{Nicolet:98534}.

A hydraulic turbine converts hydraulic energy into mechanical work. It is modeled as a variable voltage source, denoted by generator $H_{t}(\cdot)$ in Fig.~\ref{fig:EEC}. The voltage is derived by solving the so-called turbine's head characteristic, which, in Francis turbines, relates it with the turbine's rotational speed, $N$, flow $Q_t$, and guide vane opening, $y$. Kaplan turbines also include the blade angle as a second regulation mechanism and a further variable in the characteristic. A second characteristic is used to determine the turbine's torque, $T_t$, as a function of $N$, $Q_t$, and $y$. Turbine's characteristics are typically derived experimentally from the plant, see, e.g., \cite{Nicolet:98534}, and are nonlinear. The characteristics model assumes that the transient behavior of the turbine can be simulated as a succession of different steady-state operating points ("quasi-static" model). It ensures sufficient accuracy for all the flow regimes and tractable computational times \cite{Nicolet:98534}. Finally, the inertia effect of the water in the turbine is modeled through the equivalent inductance of the turbine, $L_t$.

\subsection{Linearized models of torque and head}\label{sec:linearmodels}
The equivalent circuit model in Fig.~\ref{fig:EEC} and turbine's torque expression are nonlinear because the resistance of the penstock depends on the water flow (which is a component of the model state, thus resulting in a bilinear formulation) and functions $H_t$ and $T_t$,  are nonlinear in its variables. Under the assumption that the plant's set-point and conditions do not change significantly within few control periods (several seconds), the model can be approximated by, i), considering a constant resistance of the pipe and, ii), deriving a first-order Taylor approximation of $H_t$, as proposed in \cite{cassano2021performance}. With these approximations in place and considering a linearization (typically current working point of the plant), a linearized state-space model of the circuit can be derived. For the sake of illustration, the linear model of the head, $\widetilde{H}$, reads as:
\begin{align}
\begin{aligned}
\widetilde{H}_{t}(Q_t,N,y) &\approx H_{t}(Q_{t_0}, N_0, y_0) + d^H_Q \cdot (Q_t-Q_{t_0}) + \\ & + d^H_N \cdot (N-N_0) + d^H_y \cdot (y-y_0). 
\end{aligned}\label{eq:linearhead}
\end{align}
where $d^H_Q, d^H_Q, d^H_N $ are the partial derivatives of $H_{t}(Q_t,N,y)$ calculated in the operating point. They are computed by differentiating the head characteristic numerically as:
%are defined as $c_{(h,t)_Q}, c_{(h,t)_N}$ and $c_{(h,t)_y}$, where the lower case ``h" and ``t" stands for the head and torque case.
%The c- coefficient can be founded numerically. The next example shows the methodology for the assessment of the partial derivatives of $H_t$, the same approach is used for the torque.
\begin{subequations}
\begin{align}
    & d^H_Q := \dfrac{\partial H_t}{\partial Q_t}\bigg|_{Q_{t_0}} = \dfrac{H_t(Q_{t_0}+\epsilon, \cdot)-H_t(Q_{t_0}-\epsilon, \cdot)}{2\cdot \epsilon},\\
    & d^H_N := \dfrac{\partial H_t}{\partial N}\bigg|_{N_0} = \dfrac{H_t(N_0+\epsilon, \cdot)-H_t(N_0-\epsilon, \cdot)}{2\cdot \epsilon},\\
    & d^H_y := \dfrac{\partial H_t}{\partial y}\bigg|_{y_0} = \dfrac{H_t(y_0+\epsilon, \cdot)-H_t(y_0-\epsilon, \cdot)}{2\cdot \epsilon},
\end{align}
\end{subequations}
Similar expressions hold for the linear torque, where, however, partial derivative coefficients are extracted from the torque characteristic. After introducing these approximations, the model in Fig.\ref{fig:EEC} becomes a set of ordinary differential equations, which can be converted to a discrete-time state-space (e.g., with Runge-Kutta) and suitable to be used in an MPC problem, as described next.

\subsection{MPC for fatigue reduction}\label{sec:mpc}
The principle that the MPC implements to limit penstock fatigue is that, in materials, cycles of mechanical stress below a certain value (called fatigue limit) do not count towards accumulating fatigue. In other words, alternating forces acting on a component will not significantly impact its residual service life if those forces are small enough.

By leveraging the linearized HPP presented in the former section, we can infer the head (force), and in turn the stress, in the various sections of the penstock so as to enforce a maximum limit for its variations. %This idea was firstly introduced in \cite{cassano2020reduction} and refined into an MPC problem in \cite{}, and now coupled with a BESS to 
The head at the specific portion of a penstock at time $t$, namely $h_i(t)$ (i.e., the capacitor voltage in the equivalent circuit of Fig.~\ref{fig:EEC}) is computed by solving the state-space of the linearized equivalent circuit, and then extracting the correct element from the state vector. Given with $h_i(t)$, the stress of element $i$ of an open-air penstock, $\sigma_i(t)$,  expressed in Pascal (Pa),  is then estimated as
(\cite{article}):
\begin{align}
\sigma_i(t) = h_i(t)\cdot \frac{k D}{2e} && i=1, \dots, I \label{headtostress}
\end{align}
where $D$ and $e$ are the penstock diameter and wall thickness,  respectively,  and $k = g \cdot \rho$ is the conversion factor from head in meters to pressure in Pascal, with $g$ acceleration of 
gravity and $\rho$ water density in kg/m$^3$. We impose the stress to be within these limits:
\begin{align}
   \sigma_\text{nom}  - \frac{\overline{\Delta\sigma}}{2} \le \sigma_i(t)  \le   \sigma_\text{nom} + \frac{\overline{\Delta\sigma}}{2}
   \label{eqn:stress_diseq}
\end{align}
so that the largest stress variation that can occur during a transient of stress is $\Delta\sigma$.  $\Delta\sigma$, a parameter of the problem, is the fatigue limit of the penstock and is given by the plant expert based on, e.g., its SN, or Wuhler's, curve, e.g., \cite{HADLEY2018263}.

The MPC problem consists in finding a new guide vane set point, say $y^\dagger$, that delivers a regulation duty as close as possible to the one given by the original guide vane, i.e., $y^\star$ while subject to the fatigue limit of the penstock. This aims to achieve the prescribed original regulation on a best effort basis while preserving penstock service life. The similarity criterion between $y^\star$ and $y^\dagger$ is measured in terms of squared value of their difference. Before formalizing the optimization problem of the MPC, it is worth specifying that, since the head of the penstock is dynamic (due to wave of pressure traveling within the conduits), a control action at the current time interval will impact on future values of penstock's heads. Thus, it is convenient to formulate the MPC considering a prediction horizon to ensure that the penstock constraints are respected during all the transient. The length, say $T$ in number of samples, of the prediction horizon, can be formally determined by looking at the duration of the transients or by considering the wave speed and the penstock length. As future regulation duties are not available yet, we propose to estimate them with a naive, or persistent, predictor, i.e., future $y^\star$ values match the current one; in spite of its simplicity, results show that this forecast is effective already. The optimization problem of the MPC is:
\begin{subequations}\label{eq:opt}
\begin{align}
\boldsymbol{y}^\dagger = \underset{\boldsymbol{y} \in \mathbb{R}^{T+1}}{\text{arg min}} \left\{ \sum_{\tau=t}^{t+T} \left( y(\tau) - y^\star(\tau) \right)^2 \right\} \label{eq:cost}
\end{align}
subject to guide vane limits
\begin{align}
& 0 \le y(\tau) \le 1, && \tau=t,\mydots,t+T
\end{align}
and penstock model and stress constraints:
\begin{align}
& h_i(\tau+1) =  \text{Linear penstock model (Fig.~\ref{fig:EEC})} \\ 
& \underline{h} \le h_i(\tau) \le \overline{h} \label{eq:hconstraint}
\end{align}\label{eq:mpc}\end{subequations}
for all $i$ and $\tau$. In \eqref{eq:hconstraint}, $\underline{h}, \overline{h}$ are head limits, given by combining \eqref{headtostress} and \eqref{eqn:stress_diseq}.  As typical in MPC, optimization problem \eqref{eq:mpc} is applied in a receding horizon fashion, i.e., solved at each time interval with updated information; only the first element of the decision vector $\boldsymbol{y}^\dagger$ is applied. The linear models can be recomputed when the plant's conditions change from the original linearization point. By virtue of the introduced linearized model, this optimization problem is convex and tractable. 

In summary, this MPC model offers a formal method to detect cycles which might harm the penstock. By using the filtered power set-point information together with the original power set-point and BESS' constraints of the BESS, one can then derive a power set-point of the BESS. This is the main principle of the proposed splitting policy and is described in Section \ref{sec:bess}.

\subsection{Computing the BESS power set-point} \label{sec:bess}
The intent of the proposed hybrid HPP's controller is to ensure that, following the change of HPP's operating point operated by the MPC, the BESS delivers all the missing regulating power. By doing so, the original regulation commitment of the plant is respected, and the fatigue of the penstock is reduced. In this context, a straightforward way to understand the BESS power set-point is as a difference between the power output that the HPP should have delivered with the original control set-point and the one actually implemented after the filtering action of the MPC. However, whilst the latter quantity can be accessed (e.g., from measurements possibly with some delay), the former information is not available because it is unrealized, being the original control set-point never implemented. This makes this strategy not practicable in real-life. To solve this issue, we propose to estimate the unrealized power output with an estimation model, as explained in the following.

\subsubsection{Notation}
The following notation is introduced: star ($^\star$) and dagger ($^\dagger$) superscripts refer to quantities without and with the MPC filtering action, respectively. The hat symbol ($\hat{~}$) denotes quantities that are estimated (because unobserved or unavailable). Instead, quantities without a hat refer to set-points (or measurements, when specified). Based on this notation, $P^\dagger_{hpp}$ denotes the HPP power output under MPC, $P^\dagger_{bess}$ the power set-point of the BESS, $\hat{P}^\star_{hpp}$ the estimated HPP power output without MPC control action.

\subsubsection{Modeling unobserved HPP production and calculation of the BESS set-point}
As mentioned at the beginning of this subsection, $P^\star_{hpp}$ is not available, being the control action without MPC never actuated. Thus, as a first approximation, we propose to estimate it as:
\begin{align}
    \hat{P}^\star_{hpp}(t) = \hat{T}^\star(t) \cdot \hat{\omega}^\star (t) \label{eq:estimatedpower_nl}
\end{align}
where $\hat{T}^\star$ and $\hat{\omega}^\star$ are the estimated torque and speed of the hydraulic turbine. Because the frequency (and rotor speed, properly re-scaled by the generator's polar couples) in power grids is regulated to a nominal value (e.g., 50 or 60 Hz) and we assume a small signal context for this model (i.e., model parameters can be recomputed for new conditions and variations are assumed small within a control cycle), we assumed a constant grid frequency. This yields the following approximation:
\begin{subequations}
\begin{align}
    \hat{P}^\star_{hpp}(t) = \hat{T}^\star(t) \cdot \omega_{o}. \label{eq:estimatedtorque_nl}
\end{align}
being $\omega_{o} = 2\pi \cdot f_{o}/P_p$ the rotor pulsation in mechanical radiants per second, $f_{o}$ the nominal grid frequency, and $P_p$ the polar couples of the electric generator.
It is worth highlighting that all the modeling approximations and assumptions introduced in this section for the MPC (including the linearized models, presented in the following) are validated in the results against more complete simulation models, which include swing equation and generator models.

The problem of estimating the HPP power without MPC in \eqref{eq:estimatedpower_nl} thus reduces to estimating the torque of the hydraulic turbine in \eqref{eq:estimatedtorque_nl}. To estimate the torque, we propose to use a (linearized) model from the literature, summarized for the sake of clarity in \ref{sec:linearmodels}. For convenience in the following formulation, we denote the torque estimation model with affine function $\mathcal{T}(\cdot)$. We express the estimated torque as a function of the original guide vane, $y^\star(t)$ as:
\begin{align}
\hat{T}^\star(t) = \mathcal{T}(y^\star(t), \boldsymbol{x}^\star, \boldsymbol{\rho})
\end{align}\label{eq:estimatedHPPnoMPC}\end{subequations}
where $\boldsymbol{x}^\star$ is the model state vector, and $\boldsymbol{\rho}$ a vector of model parameters (which depend on the model linearization point, too).

With the estimated HPP power as in \eqref{eq:estimatedHPPnoMPC}, the BESS set-point at time $t$, $P^\star_{bess}(t)$, could be computed as:
\begin{align}
    P^\star_{bess}(t) = \hat{P}^\star_{hpp}(t) - P^\dagger_\text{hpp}(t), \label{eq:Pbess}
\end{align}
where $P^\dagger_\text{hpp}(t)$ is the HPP power at time interval $t$. However, the output power of the plant is not known at this stage because guide vane $y^\dagger$ is being actuated and measurements are not available yet. A way to approximate this non-causal expression is to use the plant power measurement at the former time interval, namely:
\begin{align}
    P^\star_{bess}(t) = \hat{P}^\star_{hpp}(t) - P^\dagger_\text{hpp}(t-1).
\end{align}
However, depending on the refresh rate of the control action, a measurement delay could negatively impact the actuation timing, partially invalidating the principle of using a BESS, whose point of force is exactly the quick response time. To avoid this problem, and in the same spirit as done for \eqref{eq:estimatedpower_nl}-\eqref{eq:estimatedHPPnoMPC}, we propose to estimate the HPP power at the current time interval.
By using the same assumptions and notation as above, the HPP power can be estimated as the linear estimation of the torque, this time for the guide-vane computed by the MPC, $y^\dagger$. This reads as:
\begin{align}
\hat{P}^\dagger(t) = \mathcal{T}(y^\dagger(t), \boldsymbol{x}^\dagger, \boldsymbol{\rho}) \cdot \omega_{o} \label{eq:mpctorque}
\end{align}
Finally, the BESS set-point is given by combining \eqref{eq:estimatedHPPnoMPC}, \eqref{eq:Pbess}, and \eqref{eq:mpctorque}, yielding:
\begin{align}
P^\star_{bess}(t) = \left[\mathcal{T}(y^\star(t), \boldsymbol{x}^\star, \boldsymbol{\rho}) - \mathcal{T}(y^\dagger(t), \boldsymbol{x}^\dagger, \boldsymbol{\rho}) 
\right] \cdot \omega_{o}, \label{eq:bess}
\end{align}
that is the difference between the torque linear estimates for the original and filtered guide vanes ($y^\star$ and $y^\dagger$, respectively) times the nominal grid angular pulsation. 

The closed-form expression \eqref{eq:bess} can be integrated as an additional constraint in MPC's optimization problem \eqref{eq:mpc} to obtain the set-point of both HPP and BESS systems. We note that BESS's power converter constraints can be added too (see e.g. \cite{zecchino2021optimal}) without generally altering the main properties of the problem (i.e., convexity and tractability) under certain mild modeling assumptions. Despite this can be done without requiring any special effort (and should indeed be implemented in a real-time control), converter ratings are not implemented in this paper to derive insights on BESS sizing requirements, as shown and discussed for the results. The converter is assumed four-quadrant. However, only the active power is considered in this application. The reactive power (which could support voltage regulation by complementing the action of the synchronous generator exciter) is not of special interest at this stage. 

Finally, being the proposed framework for real-time control, energy constraints can be avoided under the assumption that proper energy management at a slower pace keeps the BESS state-of-charge within acceptable levels. State-of-charge-dependent power constraints should be implemented in this case to reflect different charging and discharging power capacity as a function of the state of energy.

\section{Results and Discussion}
The results shown in this section are to assess the overall regulation capacity of the hybrid power plant compared to the HPP only, the penstock's fatigue, the BESS power contribution, quantify the estimation errors introduced by the linear models, and benchmarking the splitting policy against another from the literature. Before analyzing the results, the case study and the detailed models used to simulate the HPP are described in the next two subsections.

\subsection{Case Study Specifications}
The case study refers to a 230 MW medium-head HPP. It has a net-head of 315 meters, one Francis turbine, an open-air 1'100 meter-long penstock. The whole set of model attributes used in the simulation model are summarized in Table~\ref{tab:hpp} for reproducibility. The plant is equipped with a standard governor that includes a proportional-integral (PI) regulator with a speed droop and set-point for speed-changer setting. The PI gains are determined with the Ziegler-Nicholas method. Parameters and model performance is validated by verifying the ENTSOE qualification tests for PFR \cite{landry_methodology_nodate, Test}. The regulator implements limits for the rate-of-change and magnitude of the guide vane actuator \cite{landry_methodology_nodate}. The permanent speed droop is set to 2\%. Compared to conventional speed droops for HPPs that are between 2.5\% and 5\%, we choose a lower value to reproduce future operational settings where larger support might be required from dispatchable resources. As explained in Section II, this case study is suitable because, in these plants, regulation duties (such as  primary and secondary frequency) determine pressure fluctuations inside the hydraulic conduit, possibly damaging the penstock in the long run and leading to prohibitively expensive maintenance. 

The plant simulates primary frequency regulation using a grid frequency signal from the low-inertia power system in \cite{zuo2021performance} to reproduce conditions that might happen in future grid scenarios. It is worth highlighting that the utility of the proposed model is not limited to primary frequency regulation but, more in general, to dynamics related to any change of set-points (e.g., secondary frequency control, rescheduling). For example, it was demonstrated in \cite{Dreyer2019DigitalCF} that secondary frequency regulation leads to more severe penstock damages than primary control. Inclusions of set-points from secondary frequency control will be considered in future works.

Since at this stage we are interested in evaluating the share of total power that the BESS takes as a function of the HPP's limitations, we assume that that the BESS power converter rating is large enough to accommodate any BESS set-point request. In other words, we assume that BESS constraints in the MPC splitting problem are not binding. It is to note that, since the converter power rating and BESS energy capacity are linked by the C-rate specs of the battery cells (i.e., maximum charging or discharging current [A]/battery capacity [Ah]), we also assume that the BESS energy capacity is large enough to accommodate the required power set-point. This modeling choice enables a preliminary evaluation of the BESS's sizing requirements compared to the plant's total power.

\begin{table}[!ht]
\renewcommand{\arraystretch}{1.05}
\centering
\caption{Parameters of the HPP case study}\label{tab:hpp}
\begin{tabular}{| l|c|c|c|c|c|}
\hline
\bf Parameter & \bf Unit & \bf Value \\
\hline
%\multirow{5}{*}{\rotatebox[origin=c]{90}{Parameters}}
Nominal power & MW & 230 \\
%Number of turbines & [-] & 1 \\
Nominal head & m & 315\\
Nominal discharge & m$^3$/s & 85.3\\
Nominal speed & rpm & 375\\
Nominal torque & Nm & 5.86$\times 10^6$\\
Length of penstock & m&  1'100\\
Diameter of penstock & m & 5\\
Wave speed & m/s & 1'100\\
\hline
%Three & Four\\
\end{tabular}
\end{table}

\subsection{Simulation model}
All control models are computed with the linearized models and approximations described in the Methods section. However, all control loops are closed using the nonlinear simulation models described here. This is done in order to appreciate the extent of the modeling errors in the process and assess the accuracy of the control action in a more realistic context. In the rest of this section, outputs of the nonlinear models are sometimes referred to as ground-truth values to denote they represent the benchmark values.

The HPP is modeled with the nonlinear model discussed in \ref{sec:hppmodels}. The penstock is discretized with 20 elements, sufficient to provide an accurate representation of the hydraulic transients. The turbine is modeled with the nonlinear head and torque characteristics. The power grid is modeled as an infinite bus (i.e., the grid frequency is imposed by the rest of the power system under the assumption that its size is significantly larger than the simulated plant). The generator is described with a quasi-static model that includes the swing equation (with the moment of inertia given by the generator, shaft, and hydraulic turbine masses) and power transfer modeled with the power angle approach. 
The model is simulated in MATLAB with an integration time of 4~ms. The MPC action is re-actuated each 50~ms.

The BESS is modeled as a power source. Due to the lack of mechanical time constants and extremely high ramping rates of BESS, it is assumed that BESS can implement set-points within a single control cycle (i.e., 50 ms). BESS efficiency is assumed ideal as energy losses are of limited interest for the moment.

\subsection{Results}
Simulations were performed on one day of primary frequency regulation with both the original HPP and its hybrid version, to compare them. However, the focus in this section is given to a specif time interval of the day when the penstock stress was the largest and, thus, more interesting to discuss. However, considerations equally extend to the rest of the simulated time interval, not shown here for a reason of space.

%Comparison against a  splitting policy based on a low-pass filter from the literature is also presented. Its cut-off frequency is chosen to get similar regulation performance as in the MPC.

\subsubsection{MPC set-point and head}
\begin{figure}
\includegraphics[width = 0.95\columnwidth]{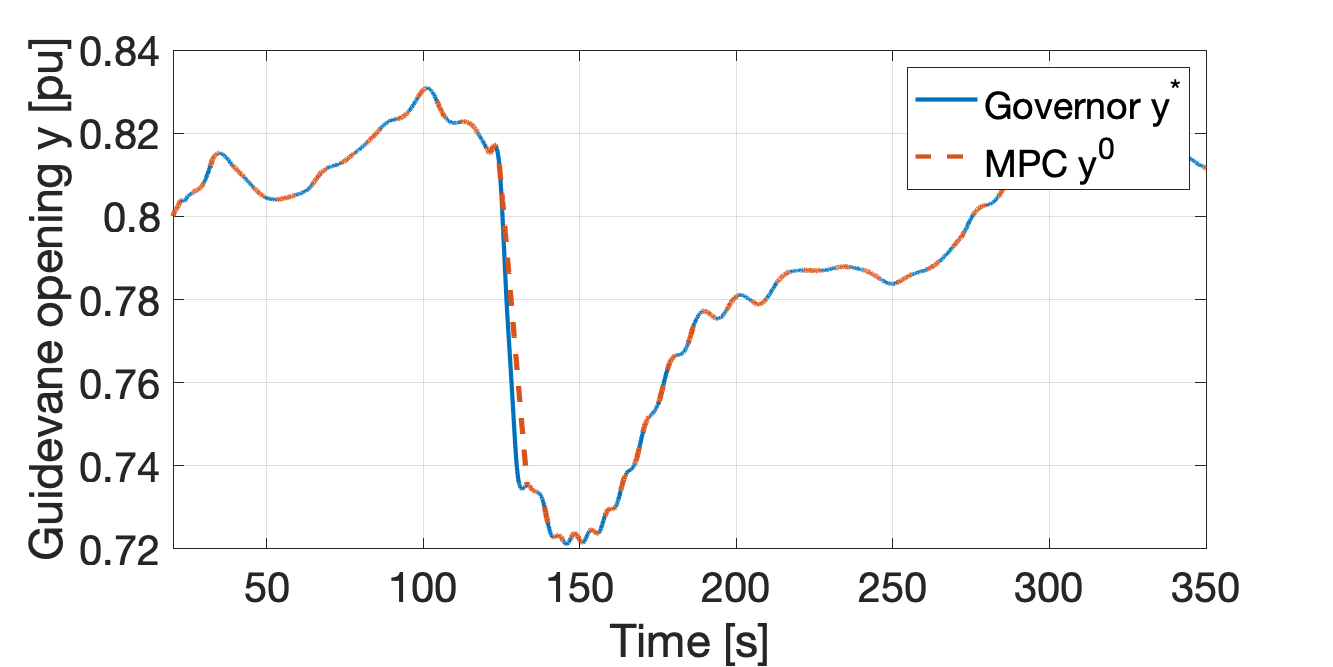} \caption{Set-point actuated by the governor and MPC.}\label{fig:gv}
\end{figure}

\begin{figure}
\includegraphics[width = 0.95\columnwidth]{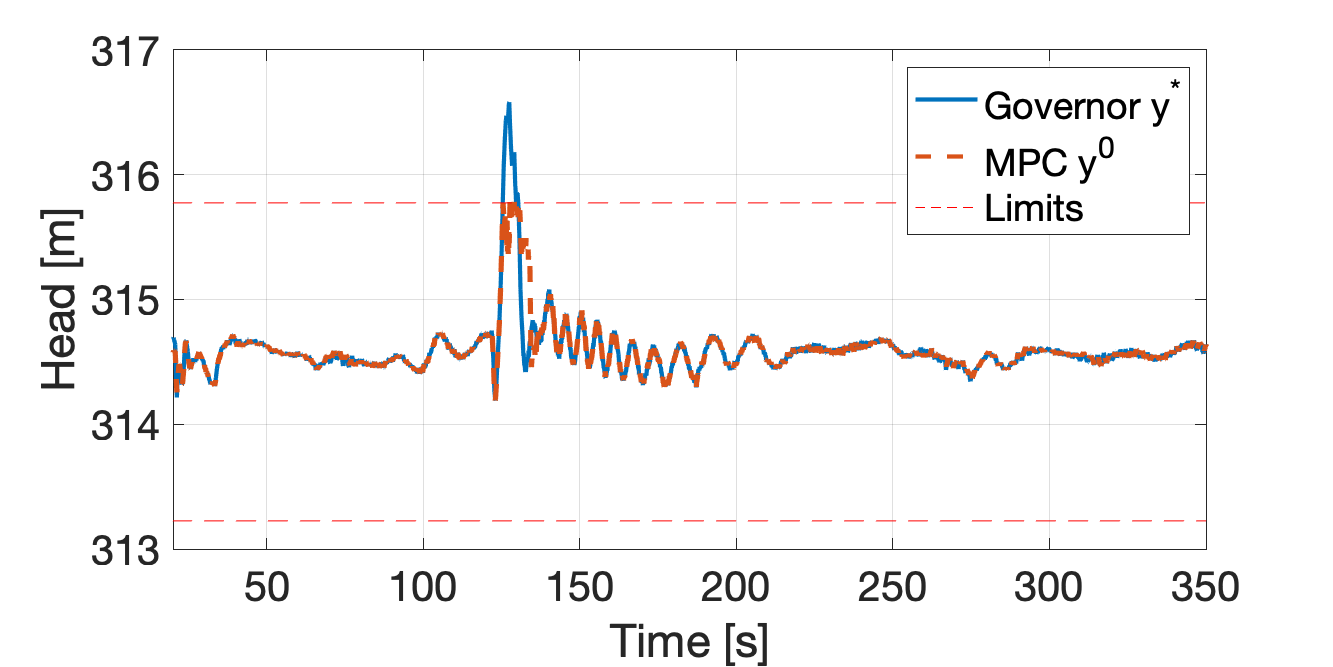}  \caption{Water heads in the critical component of the penstock from signals in Fig. ~\ref{fig:gv}. }\label{fig:heads}
\end{figure}

Fig~\ref{fig:gv} shows the HPP set-point produced by the standard governor and the one filtered by the MPC. Fig.~\ref{fig:heads} shows the head in the penstock's most critical element resulting from these control set-points and how they compare with the head limits in \eqref{eq:hconstraint}. It can be observed that: 1) the two guide vanes are identical, except when the head constraints are active. In this interval, the MPC acts as a rate limiter. This fact already gives an important indication about the low-pass filter approach proposed in the literature: excess head variations are mitigated by a filtering action that resembles the one of a rate limiter and not the one of a low-pass filter; high-frequency components alone do not seem to play a role. This will be further investigated later in a dedicated comparison. 2) The penstock head is correctly trimmed below the limit; the MPC effectively detects load cycles above a certain amplitude and reduces them. Moreover, from Fig.~\ref{fig:heads}, as the open-loop linear prediction performed by the MPC in \eqref{eq:hconstraint} is accurately below the ground-truth MPC head, we can conclude that the linear estimate is accurate enough to accomplish a successful computation of the constraint.

%t can be be seen how the head in the case of hybrid-plant respects the head limits (dashed red line) causing the damage reduction compared to the standard governor shown in Fig.~\ref{fig:RDI}. 

\subsubsection{Power output of the hybrid power plant vs the same HPP alone}
Fig.~\ref{fig:power} compares the output powers of the HPP when it provides the original amount of regulation, i.e., with guide vane $y^\star$ against the hybrid power plant (HPP + battery). This comparison is important because a requirement of the power splitting policy is to ensure that the hybrid power plant's power matches with the originally prescribed regulation.
 
Fig.~\ref{fig:power} shows that, when constraints of the head are not activated (so, away from the power dip), the hybrid power plant accurately tracks the original regulation. When constraints are activated, tracking loses accuracy progressively, with a maximum relative error over the actual PFR power of approx. 5\%; this can be explained by torque modeling error. This motivates further research on ways to reduce this error (including using recent measurements to correct the estimates) and will be considered in future work.

\subsubsection{Contribution of the BESS}
\begin{figure}
\includegraphics[width = 0.95\columnwidth]{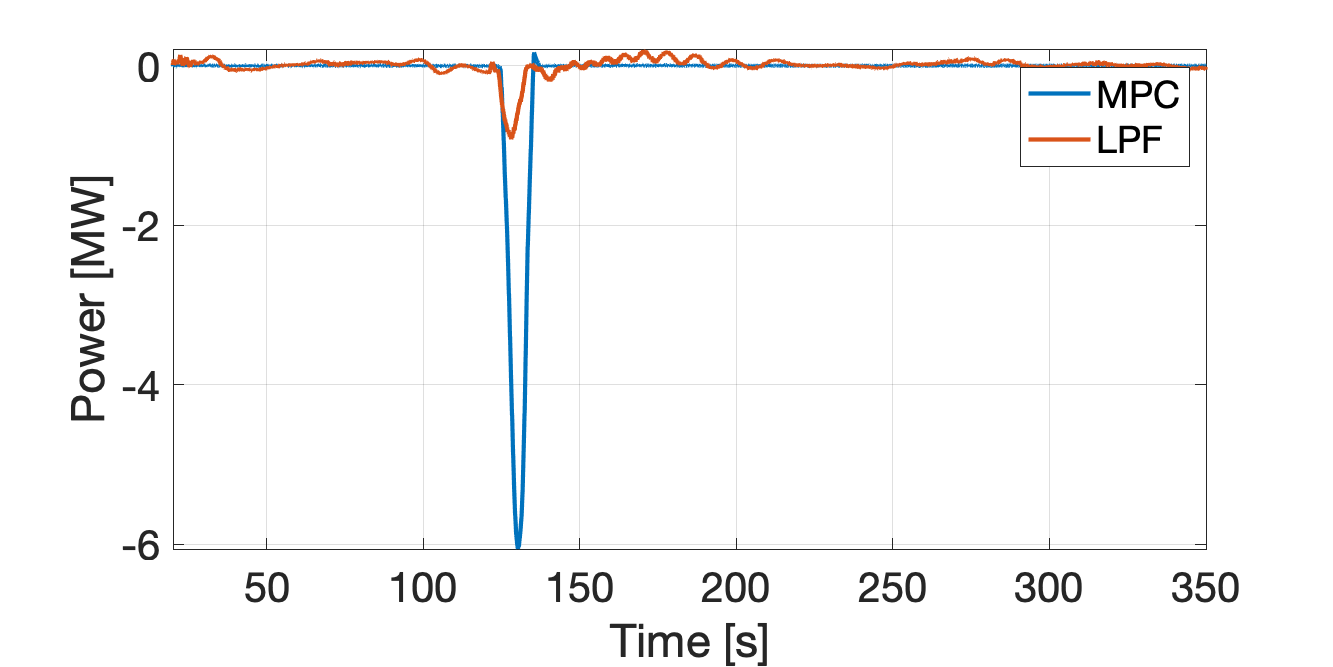} \caption{BESS power contribution in hybrid-mode with MPC and low-pass filter (LPF) control strategies.}\label{fig:battery}
\end{figure}

\begin{figure}[h!]
\includegraphics[width = 0.95\columnwidth]{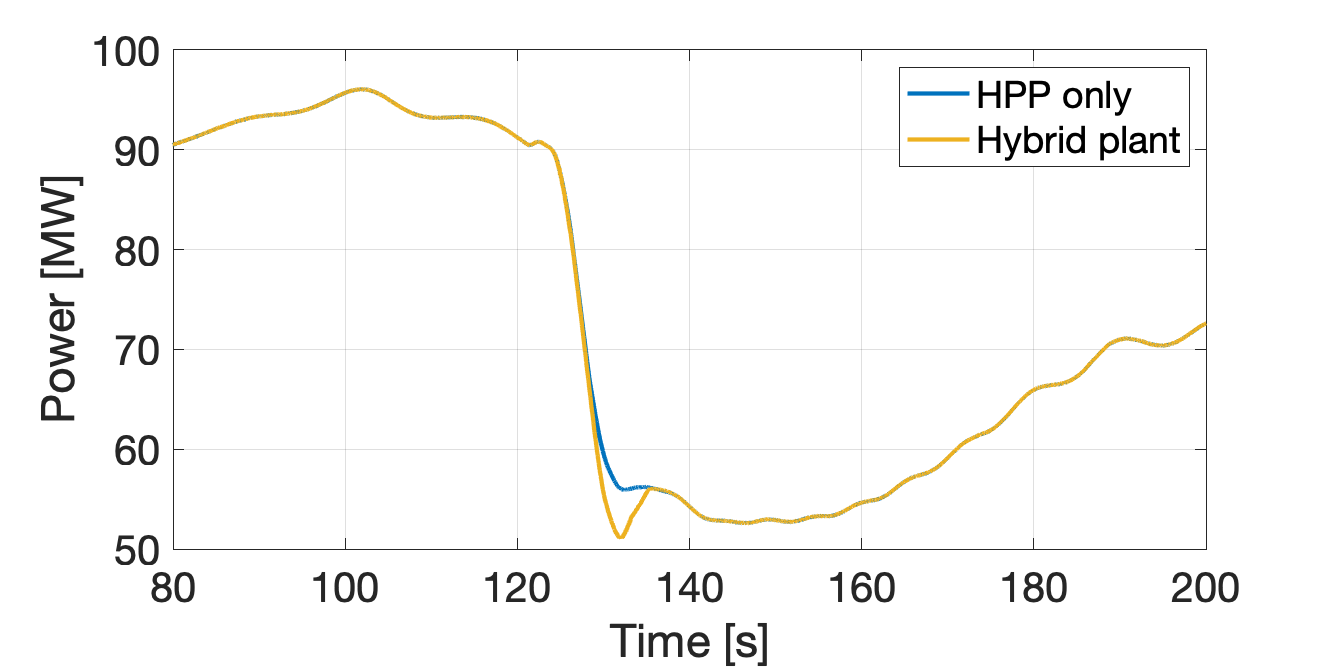}\caption{HPP output power in hybrid (yellow) and normal (red) asset with respect to the ground truth power required from PFR (blue). }\label{fig:power}
\end{figure}

Fig.~\ref{fig:battery} shows the contribution of the BESS to the hybrid power plant's output under MPC and a benchmark strategy (discussed as last).
The BESS power under MPC control compensates for the missing regulation of the HPP. It can be seen that the BESS's power demand is approx. 6~MW, less than 3\% of the plant rated power. Thus, for this specific case study and input signals, we can conclude that the power demand necessary to accomplish fatigue reduction while attaining the same tracking performance as in a plant without fatigue reduction is rather marginal. Future analysis will investigate the occurrence of these events as well as other services (secondary frequency control) as a basis to formally determining the size of a BESS. %, possibly including a tradeoff between CAPEX and missing revenues from provided regulation.

\subsubsection{Impact on penstock fatigue}

\begin{figure}
\includegraphics[width = 0.95\columnwidth]{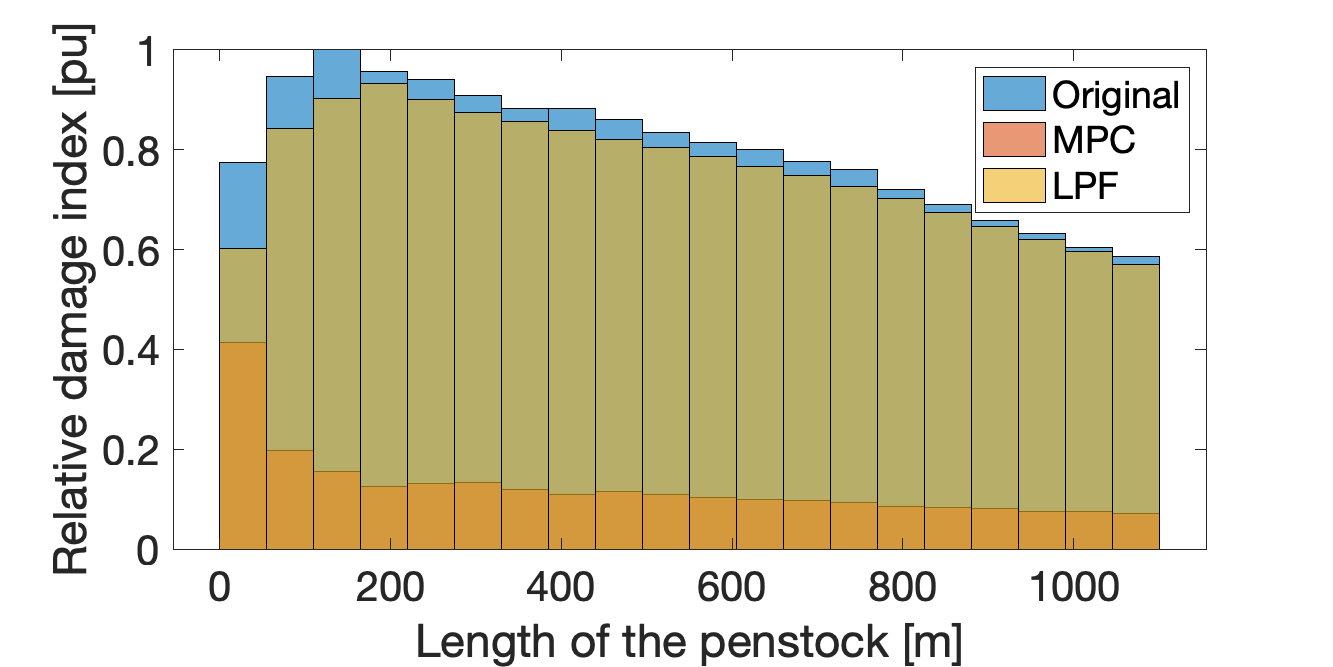} \caption{Relative cumulative damage index along the penstock with reference to the case without filtering in hybrid mode with MPC and low-pass filter. }\label{fig:RDI}
\end{figure}

It can be seen from Fig.~\ref{fig:RDI} that the penstock fatigue is reduced in the case of the hybrid power plant.  Fatigue reduction and achieving similar tracking performance as the original HPP (as shown in the former paragraph) allow us to conclude that the hybrid power plant can provide the same regulation duties as the conventional HPP for less penstock fatigue.

\subsubsection{Comparison with low-pass filter}
A comparison against a  splitting policy based on a low-pass filter (LPF) commonly proposed in the existing literature is presented here. The filter's cut-off frequency,  1.46 Hz, is chosen to get similar regulation performance (measured in terms of the correlation coefficient between powers) to when the hydropower is controlled with the MPC. In other words, the spirit here is attaining similar power output between MPC and LPF so that we can directly compare BESS contributions and effects on penstock aging in the two cases. 

BESS contributions were shown in Fig.~\ref{fig:battery}. With MPC, the BESS provides a much larger power when penstock constraints are violated. Elsewhere, the BESS is not utilized. On the other hand, LPF requires continuous action to the BESS.

Finally, the penstock damage is assessed using the procedure described in \cite{Dreyer2019DigitalCF, cassano2020reduction}. In a nutshell, it consists in calculating the head and stress in each penstock element; then, a rain-flow counting is used to count cycles at a given amplitude of stress variations; based on the SN curve (which expresses the maximum number of cycles a material can undergo at given stress) and information from the rain-flow counting, Miner's rule is applied to detect the cumulative damage index life of the component. Fig.~\ref{fig:RDI} shows the relative damage index (RDI) calculated over the 24 hours of primary frequency regulation. RDI is defined as the cumulative damage index obtained by the MPC (or LPF) divided by the cumulative damage index without any control (called base case). It can be seen that the MPC achieves to significantly reduce the damage compared to both the base case and LPF, so meeting the control objective.

\section{Conclusions}
This paper has described a model predictive control (MPC) framework for hybrid medium- and high-head hydropower (HPP) plants to determine the power set-point of the hydropower unit and the BESS. The splitting strategy relies on an explicit formulation of the mechanical loads incurred by the HPP's penstock, which can be damaged due to fatigue when providing regulation services to the grid. By removing those components conducive to excess penstock fatigue from the HPP power set-point and properly controlling the BESS, the proposed MPC was able to maintain very similar regulation levels, while significantly decreasing damages to the hydraulic conduits. Simulations were performed considering a 230 MW medium-head hydropower plant. It was observed that BESS power contribution necessary to reduce penstock fatigue was a small fraction of the rated power in this case for primary frequency control.

\bibliographystyle{IEEEtran}
\bibliography{biblio}

%\clearpage
%
%
%
%\begin{align*}
%\underset{P^\text{plant}, P^\text{battery}}{\text{min}}
%\{ \left( P^\text{plant} + P^\text{battery} - P^\text{set-point} \right)^2\}
%\end{align*}
%~~~~~~~subject to:
%\begin{align*}
%& \text{stress} = f(P^\text{plant}, \text{plant state})\\
%& \text{stress} \le \text{stress limit}\\
%& \text{Battery constraints}
%\end{align*}

\end{document}